\documentclass[10pt,aps,prl,twocolumn,amsmath,amssymb,nofootinbib,superscriptaddress]{revtex4-2}

\usepackage{xcolor}
\usepackage{mathtools} 
\usepackage{centernot} 

\usepackage[colorlinks=true,
            citecolor=red,
            linkcolor=blue,
            urlcolor=violet,
            filecolor=cyan,
            backref=false]{hyperref}

\newcommand\bs[1]{\boldsymbol{#1}}
\newcommand\dd{\mathrm{d}}
\newcommand\feq{\mathrel{\phantom{=}}}

\newcommand{\erf}{\operatorname{erf}}

\begin{document}


\title{Infinite derivative gravity resolves nonscalar curvature singularities}

\author{Ivan Kol\'a\v{r}}
\email{i.kolar@rug.nl}
\affiliation{Institute of Theoretical Physics, Faculty of Mathematics and Physics, Charles University,
V Hole\v{s}ovi\v{c}k\'ach 2, Prague 180 00, Czech Republic}
\affiliation{Van Swinderen Institute, University of Groningen, 9747 AG, Groningen, Netherlands}

\author{Tom\'a\v{s} M\'alek}
\email{malek@math.cas.cz}
\affiliation{Institute of Mathematics of the Czech Academy of Sciences, \v{Z}itn\'a 25, 115 67 Prague 1, Czech Republic}

\date{\today}

\begin{abstract}
We explicitly demonstrate that the nonlocal ghost-free ultraviolet modification of general relativity (GR) known as the infinite derivative gravity (IDG) resolves nonscalar curvature singularities in exact solutions of the full theory. We analyze exact pp-wave solutions of GR and IDG describing gravitational waves generated by null radiation. Curvature of GR and IDG solutions with the same energy-momentum tensor is compared in parallel-propagated frames along timelike and null geodesics at finite values of the affine parameter. While the GR pp-wave solution contains a physically problematic nonscalar curvature singularity at the location of the source, the curvature of its IDG counterpart is finite.
\end{abstract}

\maketitle


\section{Introduction}

Terms with higher derivatives appear in many effective descriptions of quantum gravity. They mitigate issues related to ultraviolet (UV) incompleteness of general relativity (GR) such as the existence of curvature singularities and quantum non-renormalizability \cite{Stelle:1976gc,Stelle:1977ry}. Although higher-derivative theories of finite order already regularize the linearized solutions (and their linearized curvature) to a certain order \cite{Burzilla:2020utr}, they suffer from the Ostrogradsky instability and ghosts. \textit{Infinite derivative gravity (IDG)} \cite{Krasnikov:1987yj,kuzmin1989finite,Tomboulis:1997gg,Biswas:2005qr,Modesto:2011kw,Biswas:2011ar} is a promising UV modification of GR that resolves these issues by introducing nonlocal analytic operators $\mathcal{F}(\Box)$ in the action. At the linearized level, this theory is ghost-free, singularity-free, renormalizable, and asymptotes to linearized GR in the infrared (IR). Naturally, the major open question is whether the nonlocality of IDG can resolve curvature singularities at the level of exact solutions of the full theory.

Due to the immense complexity of the field equations \cite{Biswas:2013cha}, the known exact solutions of IDG are very scarce. Putting aside the universal spacetimes \cite{Hervik:2013cla} that solve practically every gravitational theory in vacuum (up to an appropriate cosmological constant), the known exact solutions are either i) FLRW spacetimes satisfying recursive ansatz such as $\Box R=\alpha R+\beta$ \cite{Biswas:2005qr,Biswas:2010zk,Koshelev:2017tvv} or ii) pp-wave and Kundt geometries with trace-free (TF) Ricci and Weyl of types III/N \cite{Kilicarslan:2019njc,Dengiz:2020xbu,Kolar:2021rfl,Kolar:2021uiu}. Since the above recursive ansatz effectively `localizes' the field equations, the former are actually quite insensitive to nonlocality.\footnote{They are affected by $\mathcal{F}(\Box)$ only through a finite number of values, $\mathcal{F}(0)$, $\mathcal{F}(\alpha)$, and $\mathcal{F}'(\alpha)$, meaning that one can always find an equivalent local theory (with polynomial $\mathcal{F}$) that admits the same spacetimes with the same matter content.} On the other hand, the latter still retain strong dependence on the nonlocal operators $\mathcal{F}(\Box)$, so they provide better testbeds of the singularity resolution due to nonlocality.

In this Letter, we study exact pp-wave solutions of GR and IDG generated by the same energy-momentum tensor --- null radiation with the point-like distribution in the transverse 2-space and an arbitrary wave profile. In the context of IDG, these Aichelburg--Sexl-type solutions were first studied at the linearized level in \cite{Frolov:2015usa} and later promoted to the exact solutions of the full theory in \cite{Kilicarslan:2019njc}. Note that we actually consider an improved version of IDG. [In particular, our theory \eqref{eq:action} contains the Weyl term instead of the Ricci term. We also demand suppression of the propagator at both ends of the spectrum of $\bar\Box$ in \eqref{eq:ecka}, which should improve the behavior of the theory not only for short spacelike but also timelike distances.] Nevertheless, our conclusions remain valid also for the solutions from \cite{Frolov:2015usa,Kilicarslan:2019njc}. 

The goal of this Letter is to analyze the curvature of these pp-wave solutions and explicitly show the presence of curvature singularities in the GR solution and their absence in the IDG solution. Recall that \textit{curvature singularities} can be either \textit{scalar} or \textit{nonscalar}. Both of them are equally unwanted and physically problematic since they are associated with unboundedly large tidal forces experienced by an observer in a finite time. The pp-waves of type N are free from scalar curvature singularites (since all scalar invariant vanish) but they still may contain a nonscalar curvature singularity. This is defined as a point at the boundary of the manifold with diverging components of the Riemann tensor when measured in a parallel-propagated (PP) frame along a timelike or null geodesic that reaches the point with a finite value of affine parameter. In what follows, we explicitly demonstrate that the curvature in such PP frames blow up for the GR solution but remain finite for the IDG solution.

\section{Infinite derivative gravity}
Consider a general class of gravitational theories described by the Lagrangian
\begin{equation}\label{eq:action}
\begin{aligned}
    L &=\tfrac12\big[\varkappa^{-1}R-\tfrac13 R\mathcal{F}_0(\square)R+ C^{abcd} \mathcal{F}_2(\square) C_{abcd} \big]+L_{\textrm{m}}\;,
\end{aligned}
\end{equation}
where $\mathcal{F}_i$, $i=0,2$, are analytic functions of the wave operator $\Box$, $R$ is the Ricci scalar and $\bs{C}$ is the Weyl tensor.\footnote{We use the bold font for tensors and their abstract indices. The regular font is used for scalars, e.g., coordinates and tensor components.} GR is recovered by setting $\mathcal{F}_i(\Box)=0$, while IDG corresponds to non-polynomial functions $\mathcal{F}_i$, which are typically chosen by demanding specific properties of the linearized theory.

To fix a particular physically interesting IDG, let us consider the metric perturbations $\bs{\gamma}$, $\bs{g}=\bar{\bs{g}}+\bs{\gamma}$, around Minkowski spacetime $\bar{\bs{g}}$. The linearized field equations in the Landau gauge, ${\bar\nabla_{a} \gamma^{a}_{b}}=\frac{1}{4}\bar\nabla_{b} \gamma$, read
\begin{equation}\label{eq:lineqrizedfieldeq}
    -\bar\Box\mathcal{E}_2(\bar\Box){\gamma}_{ab}^{\perp} =2\varkappa {T_{ab}^{\perp}}\;,
    \quad
    -\bar\Box\mathcal{E}_0(\bar\Box) \gamma =-\tfrac{4}{3}\varkappa T\;,
\end{equation}
where the symbol $\perp$ stands for the covariant transverse-traceless projection of the symmetric rank-2 tensor and the analytic operators $\mathcal{E}_i(\bar\Box)$ are defined by
\begin{equation}
    \mathcal{E}_i(\bar\Box) =1+2\varkappa\bar\Box \mathcal{F}_i(\bar\Box)\;.
\end{equation}
Interesting choices of $\mathcal{F}_{i}(\Box)$ can be characterized by the following conditions on $\mathcal{E}_i(\bar\Box)$:
\begin{equation}\label{eq:ecka}
    \mathcal{E}_i(\bar\Box)=e^{\mathcal{A}_i(\bar\Box)}\;,
\quad \lim_{\mathclap{\bar{\Box}\to\pm\infty}}\mathcal{A}_i(\bar\Box)= +\infty\;,
\quad
    \mathcal{A}_i\left(0\right)=0\;,
\end{equation}
where $\mathcal{A}_i$ are entire functions. The first condition guarantees the absence of ghosts and other extra degrees of freedom (because the exponential function is non-zero in $\mathbb{C}$); the second condition represents the desired suppression of the propagator at short spacetime distances (i.e., in the UV); and the last condition is necessary to recover GR solutions at long spacetime distances (i.e., in the IR). 

The simplest choice satisfying \eqref{eq:ecka} is
\begin{equation}\label{eq:IDGchoice}
    \mathcal{E}_i(\bar\Box)=e^{\ell ^4 \bar\Box^2}\;.
\end{equation}
It has been demonstrated on various examples that \eqref{eq:IDGchoice} (and similar choices) regularize singular GR solutions at this linearized level. The choice \eqref{eq:IDGchoice} also fixes the non-local operators $\mathcal{F}_{i}(\Box)$ of IDG in the full theory \eqref{eq:action},
\begin{equation}\label{eq:F}
    \mathcal{F}_{i}(\Box)=\frac{e^{\ell ^4\Box^2}-1}{ 2\varkappa  \Box}\;.
\end{equation}
In what follows, we would like to demonstrate that the exact pp-wave solution of such a theory are also regular in contrast to their singular GR counterparts.

\section{Exact PP-wave solutions}

The field equations of \eqref{eq:action} reduce immensely for various subclasses of Kundt spacetimes \cite{Kolar:2021rfl,Kolar:2021uiu}. One of the simplest cases is the pp-wave metric of type N,
\begin{equation}\label{eq:metric}
\begin{aligned}
    \bs{g} = -\bs{\dd}u \vee \big(\bs{\dd}r + H\,\bs{\dd}u\big) +\bs{\dd}\rho^2 +  \rho^2 \bs{\dd}\varphi^2
\end{aligned}
\end{equation}
where $H=H(u,\rho,\varphi)$ is the only unknown function. Choosing this metric ansatz and the pure-radiation energy-momentum tensor, $\bs{T}={E}\bs{\dd}u^2$, ${E}={E}(u,\rho,\varphi)$, the field equations of the full theory reduce to
\begin{equation}\label{eq:fieldequation}
    \triangle \mathcal{E}_2(\triangle){H}=\varkappa{E}\;,
\end{equation}
where $\triangle$ is the Laplace operator on transverse 2-space ${\bs{\dd}\rho^2 +  \rho^2 \bs{\dd}\varphi^2}$. The similarity of \eqref{eq:fieldequation} with the first equation of \eqref{eq:lineqrizedfieldeq} reflects the fact that the exact solutions of the full theory within this class of metrics are also solutions of the linearized theory.\footnote{Note that this is not true for the gyraton pp-wave metrics of more general type III \cite{Kolar:2021uiu} corresponding to the spinning null matter.} This feature is purely due to the simplicity of the metric ansatz \eqref{eq:metric} --- no approximation was assumed in the derivation of \eqref{eq:fieldequation}.

Consider a source located at ${\rho=0}$ in the transverse 2-space with an arbitrary wave profile $w(u)$, ${\varkappa {E}= w(u)\delta(\rho)}$. Due to the symmetry of the source, the field equation contain only derivatives with respect to $\rho$, so the metric is characterized by one single-variable function, $h=h(\rho)$,
\begin{equation}\label{eq:hatH}
    H=w(u) h(\rho)\;.
\end{equation}
The equation \eqref{eq:fieldequation} can be solved for an arbitrary analytic choice of $\mathcal{F}_i(\Box)$ using the Fourier transform in polar coordinates that takes the form of the Hankel transform,\footnote{This integral usually contains a diverging part that is constant in $\rho$ and of no physical significance. An easy way to extract the convergent part is to replace $J_0\to\partial_{\rho}J_{0}=-sJ_1$, evaluate the $s$-integral, and calculate the primitive function in $\rho$.}
\begin{equation}\label{eq:solutionsofgenerictheory}
    h= 
    \int_0^{\infty}\!\!\!ds \frac{J_{0}(s\rho)}{-s\mathcal{E}_2\left(-s^2\right)}\;,  
\end{equation}
where $J_{0}$ is the Bessel function of the first kind.

The solutions for GR and IDG from the previous section read
\begin{equation}
\begin{aligned}
    h_{\textrm{GR}} &=
    \log\rho+C\;,
    \\
    h_{\textrm{IDG}} &=
        \tfrac{\sqrt{\pi } \rho ^2 }{2^4 \ell ^2}\, _1F_3\left(\tfrac{1}{2};1,\tfrac{3}{2},\tfrac{3}{2};\tfrac{\rho ^4}{2^8 \ell ^4}\right)
        \\
        &\feq-\tfrac{\rho ^4}{2^8 \ell ^4}\, _2F_4\left(1,1;\tfrac{3}{2},\tfrac{3}{2},2,2;\tfrac{\rho ^4}{2^8 \ell ^4}\right){+}C\;,
\end{aligned}
\end{equation}
where $C$ is a constant of no physical significance \cite{griffithspodolsky2009}. The metric components of the GR pp-wave solution diverge at the location of source ($\rho=0$) because ${h_{\textrm{GR}}\propto\log\rho}$. On the other hand, the metric components of the IDG solution are smooth at ${\rho=0}$ because $h_{\textrm{IDG}}$ is analytic as a function of $\rho^2$, which follows from analyticity of generalized hypergeometric functions $_pF_q$ and $h_{\textrm{IDG}}^{(2n+1)}(0)=0$. As we will see, these properties of $h$ are intimately related to the presence/absence of curvature singularities.

\section{Geodesics and PP frames}

Since all scalar curvature invariants vanish, we have to analyze the presence/absence of nonscalar curvature singularities. Hence, we have to investigate the geodesic motion of massive and massless test particles near $\rho=0$. For an arbitrary geodesic $\mathrm{x}(\lambda)=(u(\lambda),r(\lambda),\rho(\lambda),\varphi(\lambda))$, the corresponding tangent vector $\bs{v}$ is given by
\begin{equation}
\bs{v}=\tfrac{\bs{\mathrm{D}}\mathrm{x}}{\mathrm{d}\lambda}=u'\bs{\partial}_{u}+r'\bs{\partial}_{r}+\rho'\bs{\partial}_{\rho}+\varphi'\bs{\partial}_{\varphi}\;,
\end{equation}
where $\lambda$ is the affine parameter and $'=d/d\lambda$.

The pp-wave spacetimes possess symmetries which generate conserved quantities and simplify the integration of the geodesic equation. If we also assume that the wave-profile function is constant, ${w(u) = \text{const.}}$, the metric \eqref{eq:metric} is independent of $u$, $r$ and $\varphi$, which gives rise to three Killing vectors $\bs{\partial}_u$, $\bs{\partial}_r$, $\bs{\partial}_\varphi$ and consequently three constants of motion $c_{(u)}$, $c_{(r)}$, $c_{(\varphi)}$. In addition to that, $\bs{v}$ fulfills the normalization condition with the normalization constant $\epsilon$, hence
\begin{equation}\label{eq:COM_ppwaves}
\begin{aligned}
\bs{v}^{\flat}\cdot\bs{\partial}_{u}  &= -r' - 2 w h u' = c_{(u)}\;,
\\
\bs{v}^{\flat}\cdot\bs{\partial}_{r} &= -u' = c_{(r)}\;,
\\
\bs{v}^{\flat}\cdot\bs{\partial}_{\varphi} &= \rho^2 \varphi' = c_{(\varphi)}\;,
\\
\bs{v}^{\flat}\cdot\bs{v} &= -2 r' u' - 2 w h u'^2 + \rho'^2 + \rho^2 \varphi'^2 = \epsilon\;,
\end{aligned}
\end{equation}
where the ${\epsilon=-1}$ for timelike geodesics and ${\epsilon=0}$ for null geodesics.

Let us now we determine the PP frames along all causal geodesics ($\epsilon = -1,0$) reaching ${\rho=0}$. We start with the timelike geodesics ($\epsilon = -1$) and identify the first frame vector with the timelike tangent vector $\bs{v}$. The remaining frame vectors are chosen to satisfy the orthonormality condition, i.e., 
\begin{equation}\label{eq:ONframe}
\begin{aligned}
    \bs{e}_1 &= \bs{v}\;,  &
    \bs{e}_3 &= \tfrac{\rho'}{u'} \bs{\partial}_r + \bs{\partial}_\rho\;,
    \\
    \bs{e}_2 &= \bs{v} - \tfrac{1}{u'} \bs{\partial}_r\;,  &
    \bs{e}_4 &=  \tfrac{\rho\varphi'}{u'} \bs{\partial}_r + \tfrac{1}{\rho} \bs{\partial}_\varphi\;.
\end{aligned}
\end{equation}
The only non-vanishing components of the TF Ricci tensor $\bs{S}$ and Weyl tensor $\bs{C}$ of the metric \eqref{eq:metric} in the orthonormal frame \eqref{eq:ONframe} are given by
\begin{equation}\label{eq:SCcompONframe}
\begin{aligned}
    {S}_{11} &= {S}_{22} = {S}_{12} = w \Xi_S u'^2\;, \\
    {C}_{1313} &= {C}_{1323} = {C}_{2323} = - {C}_{1414} \\
    &= - {C}_{1424} = - {C}_{2424} = w \Xi_C u'^2\;, 
\end{aligned}
\end{equation}
where
\begin{equation}\label{eq:XiSC}
    \Xi_S = h'' + \tfrac{1}{\rho} h'\;, \quad
    \Xi_C = h'' - \tfrac{1}{\rho} h'\;.
\end{equation}
Note that the vectors $\bs{e}_3$ and $\bs{e}_4$ are PP along the geodesic with the tangent vector $\bs{e}_1$ only if ${\varphi' = 0}$; otherwise they have to be properly rotated. However, as we show bellow, $\varphi$ is constant for all geodesics passing through the source in the IDG solution.

In the case of null geodesics ($\epsilon = 0$) with the null tangent vector $\bs{v}$, we construct the corresponding complex null frame
\begin{equation}\label{eq:nullframe}
    \bs{l} = \bs{v}\;, \quad
    \bs{n} = \tfrac{1}{u'} \bs{\partial}_r\;, \quad
    \bs{m} = \tfrac{1}{\sqrt{2}} (\bs{e}_3 - i \bs{e}_4)\;,
\end{equation}
where the tangent vector $\bs{v}$ is identified with the null frame vector $\bs{l}$ and $\bs{e}_3$, $\bs{e}_4$ are defined as in \eqref{eq:ONframe}.
Similarly as for the orthonormal frame \eqref{eq:ONframe}, the null frame \eqref{eq:nullframe} is PP along $\bs{l}$ only if $\varphi'=0$.
In the standard notation of the Newman--Penrose formalism, the only non-vanishing components of the TF Ricci tensor $\bs{S}$ and Weyl tensor $\bs{C}$ of the metric \eqref{eq:metric} then read
\begin{equation}\label{eq:SCcompNullframe}
    \Phi_{22} = \tfrac{w}{2} \Xi_S u'^2\;, \quad
    \Psi_4 = \tfrac{w}{2} \Xi_C u'^2\;. 
\end{equation}

\section{GR pp-wave solution exhibits curvature singularity}

It is obvious that ${\Xi_\text{S}=0}$ for $\rho>0$ while $\Xi_\text{C}$ diverges at the location of the source, ${\rho=0}$, for the GR solution $h_{\textrm{GR}}$. Therefore also the non-vanishing components of the Weyl tensor $\bs{C}$ in PP frames along any causal geodesic reaching $\rho=0$ necessarily diverge. To prove the presence of non-scalar curvature singularity in the GR pp-wave solution, it only remains to show that the boundary points ${\rho=0}$ can be reached by at least one particular causal geodesic in a finite value of affine parameter.

\begin{figure}
    \centering
    \includegraphics[width=\columnwidth]{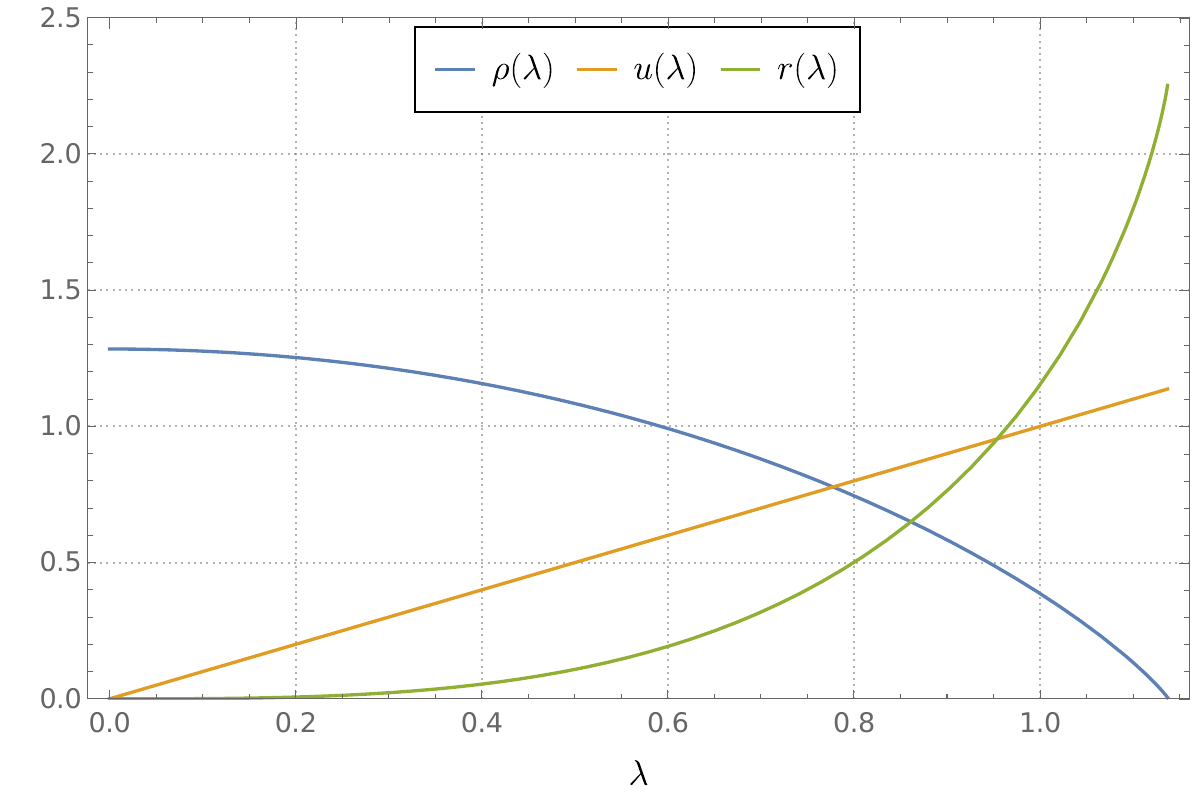}
    \caption{Coordinate description of the timelike geodesic \eqref{eq:geodGR} in the GR pp-wave solution that reaches the source (${\rho=0}$) for the finite value of ${\lambda=\lambda_0\approx 1.14}$.}
    \label{fig:geod_ppwave}
\end{figure}

An explicit timelike geodesic with vanishing angular momentum per unit mass in the transverse 2-space, ${c_{(\varphi)} = 0}$, can be obtained analytically by integration of \eqref{eq:COM_ppwaves}. Choosing the remaining constants of motion ${c_{(u)} = -1}$, ${c_{(r)} = -1}$, and the initial conditions ${u(0)=0}$, ${r(0)=0}$, ${\rho(0) = e^{1/4}}$, one gets
\begin{equation}\label{eq:geodGR}
\begin{aligned}
    u &= \lambda\;, \quad r = 2 \left( \lambda - \rho \erf^{-1} \left(\lambda /\lambda_0\right) \right)\;, \\
    \rho &= \exp\left[{1}/{4} - \left( \erf^{-1} \left(\lambda /\lambda_0\right) \right)^2 \right], \quad \varphi = \text{const.}\;,
\end{aligned}
\end{equation}
where we denoted ${\lambda_0=e^{1/4} \sqrt{\pi}/2}$. This geodesic ends up at the location of the source, ${\rho=0}$ [in particular at $(u,r,\rho,\varphi)=(\lambda_0,2\lambda_0,0,\textrm{const.})$], for the finite value of proper time, ${\lambda = \lambda_0}$. Graphs of the functions \eqref{eq:geodGR} are depicted in Fig.~\ref{fig:geod_ppwave}.

\section{IDG pp-wave solution is regular}

In the previous section, we studied only one particular geodesic as it is sufficient to show the presence of nonscalar curvature singularities. In order to verify the absence of such singularities in the IDG pp-wave solution, we have to discuss all possible causal geodesics passing through the source, $\rho=0$. 

First, we notice that $\varphi(\lambda)$ is constant along such geodesics in the IDG solution. If one assumes ${\varphi' \neq 0}$, it follows from \eqref{eq:COM_ppwaves} that ${c_{(\varphi)} \varphi' = (c_{(\varphi)}/ \rho)^2 \to + \infty}$ on the source, ${\rho = 0}$. Employing the constants of motion, $r'$ and $\rho^2$ can be eliminated in the normalization condition
\begin{equation}
    2w h_{\textrm{IDG}} c_{(r)}^2 - 2 c_{(u)} c_{(r)} + \rho'^2 + c_{(\varphi)} \varphi' = \epsilon\;.
\end{equation}
Since $h_{\textrm{IDG}}$ is bounded and $\rho'^2$ is positive, there is no way how to compensate ${c_{(\varphi)} \varphi' \to + \infty}$ on the source and therefore necessarily ${\varphi' = 0}$ implying ${c_{(\varphi)}=0}$. On the other hand, the vanishing of the constant of motion ${c_{(\varphi)}=0}$ then ensures that ${\varphi' = 0}$ also for ${\rho \neq 0}$ and we can thus conclude that $\varphi$ is constant along any geodesic passing through the source.

As an immediate consequence of this statement, the orthonormal frame \eqref{eq:ONframe} and the null frame \eqref{eq:nullframe} are PP along the corresponding geodesics. It turns out that the components of the curvature tensors ($\bs{S}$ and $\bs{C}$) in PP frames along any timelike and null geodesics passing through the source are given exactly by \eqref{eq:SCcompONframe} and \eqref{eq:SCcompNullframe}, respectively. These components are finite everywhere since $\Xi_S$, $\Xi_C$ are bounded and $u'=-c_{(r)}$ for IDG pp-waves solution $h_{\textrm{IDG}}$.


\section{Conclusions}

We have shown that IDG can resolve nonscalar curvature singularities in exact pp-wave solutions. Thus, the nonlocality of IDG can prevent arbitrarily large tidal forces that would be experienced by an observer in finite time in these spacetimes. This is the first explicit exact demonstration of the singularity resolution due to nonlocality where the curvature of the exact GR and IDG solutions with the same matter content were compared side by side. These results were achieved by analyzing components of the curvature tensors in PP-frames along timelike and null geodesics reaching the source with the finite value of affine parameter.

We believe that our study will stimulate further search for exact solutions of IDG and examination of the presence/absence of (scalar or nonscalar) curvature singularities due to nonlocality. For example, a similar calculation can be repeated for gyraton solutions generated by spinning null matter \cite{Kolar:2021uiu} or Kundt geometries with non-zero cosmological constant \cite{Dengiz:2020xbu,Kolar:2021rfl}. The former would be particularly interesting because the exact IDG gyratons \cite{Kolar:2021uiu} differ from the linearized IDG gyratons \cite{Boos:2020ccj}. A careful analysis of spacetime singularities in nonlocal gravitational theories (and the associated matter distribution) should also reduce the lack of clarity on this topic in the current literature \cite{Buoninfante:2018rlq}.


\section*{Acknowledgements}

I.K. was supported by Netherlands Organization for Scientific research (NWO) grant no. 680-91-119 and Primus grant PRIMUS/23/SCI/005 from Charles University. T.M. acknowledges the support of the Czech Academy of Sciences (RVO 67985840) and the Czech Science Foundation GA\v{C}R grant no. GA19-09659S.



\bibliographystyle{apsrev4-2}

%

\end{document}